\documentclass[a4paper]{jpconf}
\usepackage{graphicx}
\begin{document}
\title{Fine structure of the isoscalar giant monopole resonance in $^{48}$Ca}

\author{S.D. Olorunfunmi, I.T.~Usman, J.~Carter, L.~Pellegri, P.T.~Molema, E.~Sideras--Haddad}

\address{School of Physics, University of the Witwatersrand, Johannesburg 2050, South Africa}

\author{R. Neveling, F.D. Smit, P.~Adsley, L.M.~Donaldson, L.~Pellegri, G.F.~Steyn}

\address{Department of Subatomic Physics, iThemba LABS, Somerset West 7129, South Africa}

\author{P. von Neumann-Cosel, N.~Pietralla }

\address{Institute f\"{u}r Kernphysik, Technische Universit\"{a}t Darmstadt, D-64289 Darmstadt, Germany}

\author{N.N. Arsenyev}
\address{Bogoliubov Laboratory of Theoretical Physics, Joint Institute for Nuclear Research, 141980 Dubna, Moscow region, Russia}

\author{P. Papka, K.C.W.~Li, J.W.~Br\"{u}mmer}
\address{Department of Physics, University of Stellenbosch, Matieland 7062, South Africa.}

\author{G.G.~O'Neill, V.~Pesudo, D.J.~Mar$\acute{i}$n--L$\acute{a}$mbarri}
\address{Department of Physics, University of the Western Cape, Private Bag X17, Bellville 7535, South Africa}

\author{H. Fujita, A.~Tamii}
\address{Research Center for Nuclear Physics, Osaka University, Ibaraki, Osaka 650-0047, Japan}


\begin{abstract}
Experiments investigating the fine structure of the IsoScalar Giant Monopole Resonance (ISGMR) of $^{48}$Ca were carried out with a 200 MeV alpha inelastic-scattering reaction, using the high energy-resolution capability and the zero-degree setup at the K600 magnetic spectrometer of iThemba LABS, Cape Town, South Africa. Considerable fine structure is observed in the energy region of the ISGMR. Characteristic energy scales are extracted from the experimental data by means of a wavelet analysis and compared with the state-of-the-art theoretical calculations within a Skyrme-RPA (random phase approximation) approach using the finite-rank separable approximation with the inclusion of phonon-phonon coupling (PPC). Good agreement was observed between the experimental data and the theoretical predictions.


\end{abstract}

\section{Introduction}
Nuclear giant resonances are small-amplitude, high-frequency, simple collective modes of excitation. They are considered to be the fastest vibrations of any known many-body system, and are characterized by excitation energies above the particle-emission threshold and broad widths greater than 2.0 MeV \cite{Har02}. The study of the IsoScalar Giant Monopole Resonance (ISGMR) is important since knowledge of its excitation energy provides information relevant to the nuclear-matter incompressibility \cite{Bor75, Bla80}. This is of astrophysical importance as the incompressibility of the nuclear matter is used to constrain the nuclear equation of state (EOS). Furthermore, various studies on giant resonances via high energy-resolution inelastic particle-scattering experiments have established fine structure as a generic phenomenon of nuclei irrespective of the mode \cite{Kal81, Kal06}. The observed fine structure provides valuable information about factors that contribute to the decay of giant resonances. Specifically, at the iThemba Laboratory for Accelerator Based Sciences (iThemba LABS), a series of experimental studies using  proton scattering has established the systematic observation of fine structure of the ISGQR and IVGDR \cite{Shev08, Usm11, Jin18, Kur18, Lat19}. The present study seeks to answer the question regarding similar observations for the ISGMR in calcium isotopes. The work presented here is a very preliminary result from ongoing work at iThemba LABS to confirm the fine structure phenomenon for the ISGMR.
A viable way of understanding the origin and nature of fine structure is to extract the characteristic energy scales from the experimental data and compare them with the most suitable theoretical models. A quantitative description of the fine structure in terms of characteristic scales is derived using wavelet techniques. The present work employs the microscopic model to investigate the nature of fine structure in the ISGMR for the case of $^{48}$Ca. Following the basic ideas of the quasiparticle-phonon model (QPM) \cite{Sol92}, the model Hamiltonian has been generalized for the phonon-phonon coupling (PPC) based on an effective Skyrme interaction \cite{Sev04}. This model takes into account the coupling between the one- and two-phonon terms in the $0^{+}$ wave functions.


\section{Experiments}
\begin{figure}[h]
\begin{center}
\includegraphics[width=25pc]{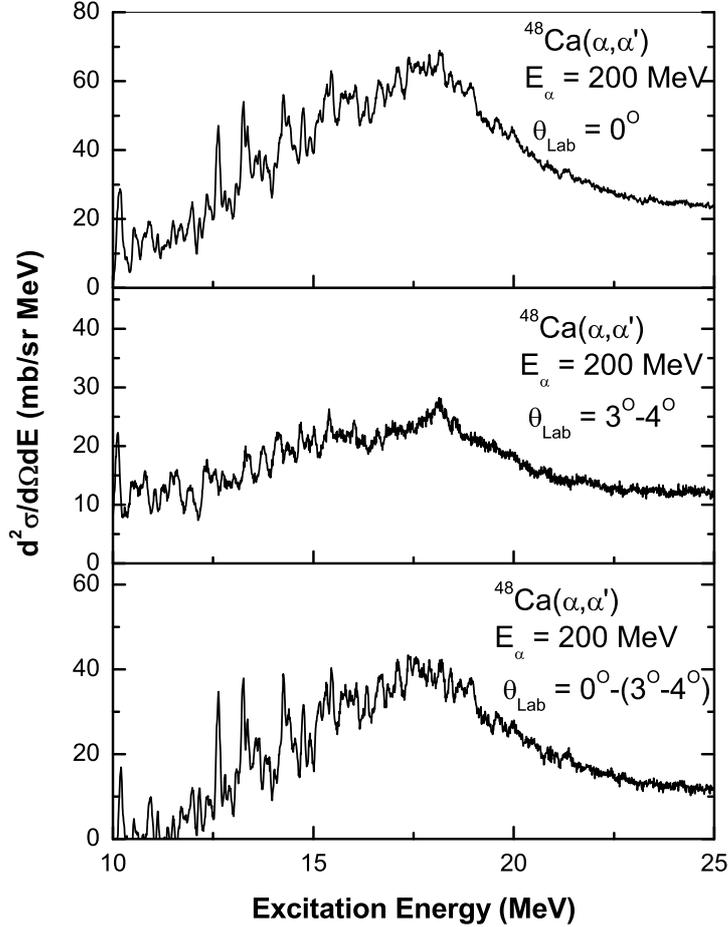}\hspace{2pc}%
\caption{\label{fig:nm3}Upper panel: Experimental double differential cross-sections for the $^{48}$Ca($\alpha,\alpha'$) reactions at 0$^{\circ}$ with \textit{E$_{\alpha}$} = 200 MeV. Middle panel: Experimental double differential cross-sections for the 3-4$^{\circ}$ angle cut. Lower panel: Difference of spectra obtained as described in the text.}
\end{center}
\end{figure}

Inelastic scattering of 200 MeV alpha particles on $^{48}$Ca was measured at the Separated Sector Cyclotron (SSC) facility of iThemba LABS in Cape Town, South Africa. A self-supporting foil of highly enriched $^{48}$Ca target with areal density of 1.43 mg/cm$^2$ was used. Inelastically scattered  $\alpha$--particles were momentum analyzed with the high energy-resolution K600 magnetic spectrometer. The horizontal and vertical positions of the focal plane were determine using a detector system consisting of two Vertical Drift Chambers (VDCs) and a pair of plastic scintillation detectors. Data were taken with the K600 spectrometer at 0$^{\circ}$ (0$^{\circ}$-1.91$^{\circ}$ angle acceptance) and 4$^{\circ}$ (2$^{\circ}$-6$^{\circ}$ angle acceptance) laboratory angles. A full description of the experimental setup and the measurement techniques is given in Ref. \cite{Nev11}. 
Experimental spectra obtained for $^{48}$Ca are shown in Fig. \ref{fig:nm3}. The upper panel of Fig.\ref{fig:nm3} shows the double differential cross-section at 0$^{\circ}$, where the ISGMR strength is maximal, and the middle panel shows the double differential cross-section for the 3$^{\circ}$-4$^{\circ}$ angle cut where the first minimum of the ISGMR angular distribution is expected to lie. The difference-spectrum (in the lower panel of Fig.\ref{fig:nm3}), obtained by subtracting the 3$^{\circ}$-4$^{\circ}$ spectrum from the 0$^{\circ}$ spectrum, has been established to essentially represents only the ISGMR strength \cite{Bra87,You97}. Pronounced fine structure of the ISGMR can be observed in the excitation energy range of 10--25 MeV. 

\section{Theoretical calculation}
Giant Resonances (GRs) are damped with a width of a few MeV, and microscopic approaches are employed to better understand the various factors that contribute to the total width and strength of the resonances. This is achieved by using suitable microscopic models to interpret the characteristic energy scales extracted from the experimental results.

\begin{figure}[h]
\begin{center}
\includegraphics[width=20pc]{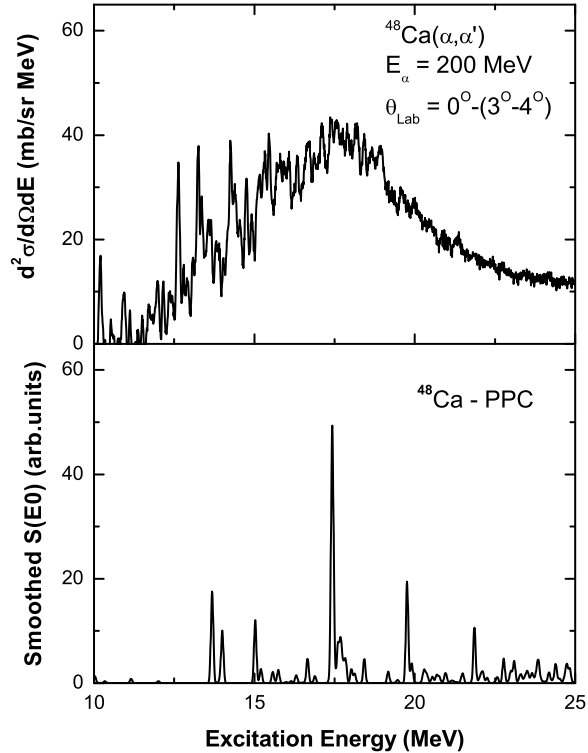}
\caption{\label{fig:nm1}Upper panel: Experimental difference-spectrum of the $^{48}$Ca($\alpha,\alpha'$) reaction. Lower panel: the \textit{E}0 strength function is calculated with the inclusion of PPC.} 
\end{center}
\end{figure}

The pronounced fine structure observed in the experimental results shown in Fig. \ref{fig:nm3} has been analyzed by calculating the \textit{E}0 strength function for $^{48}$Ca using the PPC formalism. The method has been discussed in detail in \cite{Sev04,Ars17}, however we recall it for completeness. The Hartree-Fock-BCS (HF-BCS) calculations are performed by using the SLy5 Skyrme interaction in the particle-hole channel. Making use of the finite rank separable approximation for the residual interaction enables one to perform the random phase approximation (RPA) in very large 1p-1h spaces (see Ref. \cite{Sev08} for more details). To take into account the effects of the PPC, we follow the basic QPM ideas \cite{Sol92}. We construct the wave functions of excited states as a linear combination of one- and two-RPA phonon configurations. To construct the wave functions of the $0^{+}$ states, in the present study we take into account all two-phonon configurations below 30 MeV that are built from the phonons with different multipoles $\lambda^{\pi}=0^{+},1^{-},2^{+},3^{-},4^{+}$ and $5^{-}$ coupled to $0^{+}$. The calculated PPC \textit{E}0 strength was smoothed with a Gaussian function with a width comparable to the energy resolution of the experiment, i.e. FWHM = 85 keV. This is to allow for a direct comparison between experiment and theory as shown in Fig. \ref{fig:nm1}. The lower panel of Fig. \ref{fig:nm1} shows the result of the theoretical calculation. It can be seen from Fig. \ref{fig:nm1} that the PPC results compare reasonably well with the experimental results with regards to the centroid energy and width exhibited. It should be noted that the experimental spectrum does not represent the \textit{E}0 strength, but the double differential cross-section.


\section{Wavelet analysis results and discussion}

\begin{figure}
\begin{center}
\includegraphics[width=30pc]{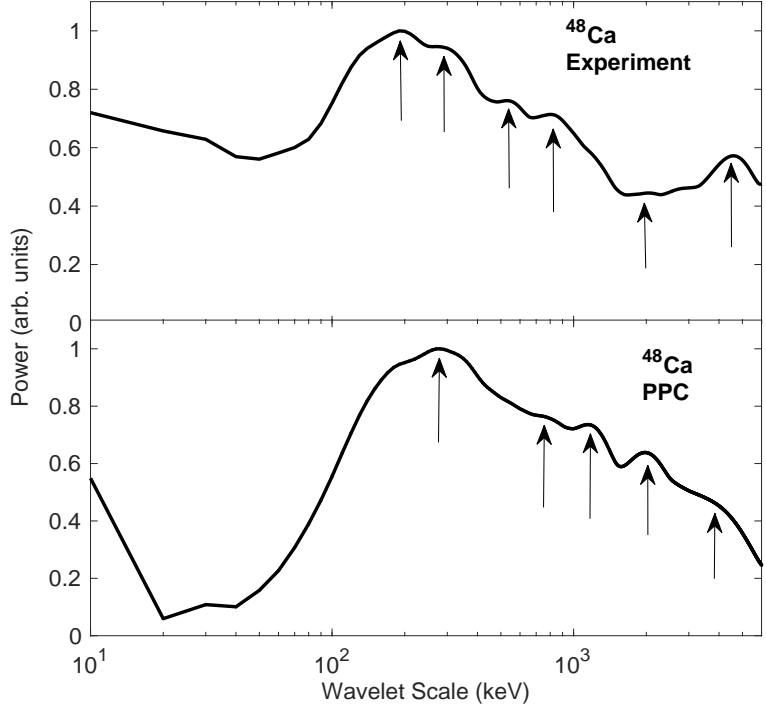}
\caption{\label{fig:nm2}Wavelet power spectra showing energy scales in experimental and theoretical spectra for the $^{48}$Ca ISGMR. The arrows indicate characteristic scales.}
\end{center}
\end{figure}

A Continuous Wavelet Transform (CWT) analysis technique \cite{Shev08} was used to obtain characteristic energy scales from both the experimentally-measured excitation energy spectrum for $^{48}$Ca($\alpha,\alpha'$) and the theoretically calculated \textit{E}0 strength function. In order to obtain these characteristic scales, a power spectrum of the wavelet signal was applied on the scale axis. The results are shown in Fig. \ref{fig:nm2}. The values of the extracted scales and their classes are presented in Table \ref{tb:sc}. Class I contains scales below 300 keV. Intermediate-energy scales of the order of several hundreds of keV belong in Class II, while Class III contains scales above 1 MeV. From the power spectra in Fig. \ref{fig:nm2} and Table \ref{tb:sc}, it can be observed that all the three classes of scales are present in both the experimental data and the PPC calculations.

\begin{table}[h]
\caption{\label{ex}Comparison of scales (in keV) observed in both the experimental data and the theoretical calculations.}
\begin{center}
\begin{tabular}{llll}
\br
$^{48}$Ca &Class I&Class II&Class III\\
\mr
Experiment &190 280 & 540 820&1050 1960 4450\\
PPC & ----- 280&----- 780&1150 1980 3960 \\
\br
\end{tabular}
\label{tb:sc}
\end{center}
\end{table}

\section{Conclusions}

Inelalstic $\alpha$-particle scattering measurements, with \textit{E$_{\alpha}$} = 200 MeV and $\theta_{Lab}$ = 0$^{\circ}$ and 4$^{\circ}$, were carried out on $^{48}$Ca. Considerable fine structure was observed in the excitation energy region of the ISGMR. Characteristic energy scales were extracted in the resonance region using the CWT technique for both the experimental data and the theoretical results. The analysis revealed three classes of scales. A good agreement between the experimental data and the theoretical predictions in all classes of scales was achieved. Further comparison of the experimental data with some other microscopic models might be necessary in order to make more definite statements on the dominant decay mechanism that contributes to the total width of the ISGMR. A study of the evolution of the ISGMR as a function of neutron excess in calcium isotopes is underway.

\end{document}